 \documentstyle[twoside,psfig]{elsart}  
\begin{document}

\begin{frontmatter}

\title	{Waiting-times and returns in high-frequency financial data:
an empirical study}

\author[Genova] {Marco Raberto},
\author[Alessandria]{Enrico Scalas},
\author[Bologna] {Francesco Mainardi\thanksref{mail*}}
\address[Genova]  {Dipartimento di Ingegneria Biofisica ed Elettronica,
			Universit\`a di Genova, \\
		via dell'Opera Pia 11a,  I--16145 Genova, Italy}
\address[Alessandria]	 {Dipartimento di Scienze e Tecnologie Avanzate,
			Universit\`a del \\ Piemonte Orientale,
			via Cavour 84,	I--15100 Alessandria, Italy} 
\address[Bologna] {Dipartimento di Fisica, Universit\`a di Bologna and
		  INFN Sezione di Bologna, \\
		  via Irnerio 46, I--40126  Bologna, Italy}
\thanks[mail*]{Corresponding author.
 E-mail: {\tt mainardi@bo.infn.it}
 URL: {\tt www.fracalmo.org}}

\begin{abstract}
In financial markets, not only prices and
returns can be considered as random variables,
but also the waiting time between two transactions
varies randomly. In the following, we analyse
the statistical properties of General Electric stock
prices, traded at NYSE, in October 1999.
These properties are critically revised  in the
framework of theoretical predictions based on a
continuous-time random walk model.
\end{abstract}


\end{frontmatter}

\def\eg{{e.g.}\ } \def\ie{{i.e.}\ }  
\def\sg{\hbox{sign}\,}
\def\sgn{\hbox{sign}\,}
\def\sign{\hbox{sign}\,}
\def\e{\hbox{e}}
\def\exp{\hbox{exp}}
\def\ds{\displaystyle}
\def\dis{\displaystyle}
\def\q{\quad}	 \def\qq{\qquad}
\def\lan{\langle}\def\ran{\rangle}
\def\l{\left} \def\r{\right}
\def\lra{\Longleftrightarrow}
\def\arg{\hbox{\rm arg}}
\def\d{\partial}
 \def\dr{\partial r}  \def\dt{\partial t}
\def\dx{\partial x}   \def\dy{\partial y}  \def\dz{\partial z}
\def\rec#1{{1\over{#1}}}
\def\log{\hbox{\rm log}\,}
\def\erf{\hbox{\rm erf}\,}     \def\erfc{\hbox{\rm erfc}\,}
\def\FT{\hbox{{F}}\,}  \def\F{\hbox{{F}}\,}  
\def\LT{\hbox{{L}}\,}	\def\L{\hbox{{L}}\,} 
\def\NN{\hbox{\bf N}}
\def\RR{\hbox{\bf R}}
\def\CC{\hbox{\bf C}}
\def\ZZ{\hbox{\bf Z}}
\def\II{\hbox{\bf I}}


\section{Introduction}

In financial markets, waiting times between two consecutive
transactions vary in a stochastic fashion.
In 1973, \cite{Clark 73} Peter Clark wrote:
``Instead of indexing [time] by the integers 0,1,2,$\ldots$, the [price]
process could be indexed by a set of numbers $t_{1}$, $t_{2}$, $t_{3}$,
$\ldots$, where these numbers are themselves a realization
of a stochastic process (with positive increments, so that
$t_{1} < t_{2} <t_{3} < \ldots $).''

From this point of view
the continuous time random walk (CTRW) model
of Montroll and
Weiss \cite{Montroll 65} (see also Refs.
\cite{Montroll 79,Montroll 84,Weiss 94b})
can provide a phenomenological description of tick-by-tick dynamics
in financial markets \cite{SGM 00,MRGS 00,GMSR 01}.

Actually in CTRWs, two random variables are used: jumps
$\xi_n = x(t_{n+1})- x(t_n)$ and waiting times
$\tau_n = t_{n+1} - t_n$. In the financial interpretation of
CTRWs, $x$ represents a log-price and $\xi$ a log-return
\cite{SGM 00,MRGS 00,GMSR 01} (see also \cite{Parkinson 77}).
The physicist can think of $x$ as the position of a  random walker
performing discrete jumps in one dimension at  randomly
distributed instants.  Based on \cite{MRGS 00}
the evolution equation for $p(x,t)$, the probability of occurrence of the
log-price $x$ at time $t$, or
of finding the random walker at position $x$ at time instant $t\,, $
can be written,  assuming the initial condition $p(x,0) = \delta (x) $
(\ie the walker is initially at the origin $x=0),$
$$   p(x,t) =  \delta (x)\, \Psi(t) +
   \int_0^t
 \int_{-\infty}^{+\infty}  \varphi (x-x',t-t')\, p(x',t')\, dx'\,dt'
 \,, \eqno(1.1) $$
where
$\Psi(t)$ is the {\it survival probability}
and
$\varphi(\xi, \tau)$, is the
{\em joint probability density}
of jumps $\xi_{n} = x(t_{n+1}) - x(t_{n})$ and of waiting times
$\tau_n = t_{n+1} - t_{n}$.
Relevant quantities are the two (marginal) probability density functions
($pdf$'s)  defined as
$ \, \lambda (\xi) :=  \int_0^\infty
 \varphi (\xi, \tau)  \,d \tau\,,$
$\, \psi(\tau ):= \int_{-\infty}^{+\infty}
 \varphi (\xi, \tau) \,d \xi \,, $
and called {\it jump pdf} and {\it waiting-time pdf}, respectively.
If one assumes that  the {\it jump} $pdf$
$ \lambda(\xi )$   is independent of  the
{\it waiting-time} $pdf$ $\psi(\tau)\,,$
 we have the so-called "decoupling" which leads to
the  factorisation
$    \varphi(\xi ,  \tau ) = \lambda   (\xi ) \, \psi ( \tau )\,.$

The Eq. (1.1) is  the most general {\it master equation}
of the CTRW, usually derived in the Fourier-Laplace domain.
The simplified form under the hypothesis of "decoupling" 
is reported in \cite{MRGS 00}. 

The probability that a given inter-step interval is greater or equal
to $\tau $ is $\Psi(\tau)\,, $
which is defined in terms of $\psi(\tau )$ by
  $$ \Psi(\tau) =\int_\tau ^\infty \psi(t')\, dt'
  = 1- \int_0^\tau  \psi(t')\, dt'\,, \q
 \psi(\tau ) = - {d \over d\tau} \Psi(\tau)\,. \eqno(1.2)$$
We note that  $\int_0^\tau  \psi(t')\, dt'\,$ represents the
probability  that at
least one  step is taken at some instant in the interval $[0,\tau) $,
hence	$\Psi(\tau )\,$   is the probability  that the diffusing quantity
$x$ does not change value during the time interval of duration $\tau $
after a jump.
We also note, recalling that $t_0=0\,,$
that $\Psi(t)\,$
is the {\it survival probability}
until time instant $t$ at the initial
position $x_0=0\,. $

A relevant choice for the survival probability  is given by the
Mittag-Leffler function of order $\beta $
($0<\beta <1$),
which leads to a time-fractional
master equation as shown in \cite{MRGS 00} (see also
\cite{HilferAnton 95,Saichev 97}).
For reader's convenience hereafter we recall the main properties
of this transcendental function useful for our purposes.
From its definition  valid for any $\beta >0\,$:
$$ \Psi(\tau ) = E_\beta \l[-(\tau /\tau _0)^\beta \r] :=
 \sum_{n=0}^{\infty}
  (-1)^n {(\tau /\tau _0)^ {\beta n}\over\Gamma(\beta\,n+1)}
\,,\q \beta > 0\,,\eqno(1.3)$$
one recognises that the Mittag-Leffler function
generalises the simple exponential
function (recovered for $\beta=1$) 
and, if $0<\beta<1\,,$   it interpolates 
on the positive real axis a stretched exponential 
and a power law
according to
$$
E_\beta \l[-(\tau /\tau _0)^\beta \r] \sim    \cases{
     \exp \l[ -(\tau /\tau _0)^\beta /\Gamma(1+\beta)\r], 
      & $\tau/\tau_0 \to 0^+$,\cr
       (\tau /\tau _0)^{-\beta}/ \Gamma(1-\beta),
  & $\tau/\tau_0 \to \infty$,}
\; 0< \beta <1.
     \eqno(1.4) $$
For more information on the Mittag-Leffler function see
\eg \cite{Erdelyi HTF,GorMai CISM97,MaiGor 00}.

The purpose of this paper is to investigate some
statistical properties of the random variables
$\xi$ and $\tau$ in financial markets.
This study is limited to a specific equity of a given market in
a definite period. Therefore caution is necessary and our results
cannot be arbitrarily generalized. In particular,
the reader will learn about General Electric stock
prices, traded at NYSE, in October 1999. This preliminary
presentation is part of a broader project aimed at studying the
behaviour of all Dow-Jones-Industrial-Average stocks
during that month.

\section{Empirical analysis}

In Fig. 1,
a scatter plot is presented for waiting times $\tau_n$ as a
function of the corresponding log-return $\xi_n$.
By means of a contingency table analysis \cite{BL 90},
we have studied the independence of the two stochastic variables.
A direct inspection of Fig. 1 shows that
for large values of log-returns waiting times tend to be shorter.
This indicates a possible correlation.
Actually, a hypothesis test has been performed
on the empirical joint
$pdf$ $\varphi (\xi ,\tau )\,. $
According to the contingency table presented in Tab. 1,
the two random variables cannot be considered independent.
The null hypothesis of independence can be rejected with a
significance level of 1\%.


In Fig. 2, an estimate of the autocorrelation function
for the absolute value of log-returns is plotted.
We have used the following estimator
\cite{Marple 87} 

$$
C(m)=\frac{1}{N-m}
\sum_{n=0}^{N-m-1}(|\xi_{n+m}|-\overline{|\xi|})(|\xi_{n}|-\overline{|\xi}|),
\eqno(2.1)$$
where $N$ is the total number of points ($N=55559$) and
$
\overline{|\xi|} = \frac{1}{N} \sum_{n=0}^{N-1} |\xi_n|\,.
$
The inset shows the time series of
the absolute values as a function of the tick $n$.

Due to scale persistence,
the autocorrelation function
follows a power-law decay with a slope of $-0.76$.
The autocorrelation is over the noise
level (conventionally $3/\sqrt{N}$) for a lag between 20 and 30 ticks,
corresponding to an average time of $250 s$.
Therefore, within that time scale,
it is not safe to assume that the log-returns themselves,
$\xi_n$, are independent variables.
These are well-known stylised fact for tick-by-tick financial
time series, see \eg  \cite{RamaCont 99,MantegnaStanley 00,GPLAGS 00}.

In Fig. 3,
the autocorrelation function is shown for waiting-times $\tau_n$.
As above, the inset shows the time series itself.
Waiting times between trades are inherently positive random variables.
For the GE stock in October 1999,
there is a marked seasonality of waiting times with a 1-day period
(nearly $3,000$ trades).
Inspection of the series shows that the trading activity is
slower in the middle of the day.
The seasonality is outlined 
by the periodic behaviour of the autocorrelation estimate,
with periodicity above the conventional noise band.

In recent times, several efforts have been devoted
to find   
a suitable {\it measure} of time,
in order to discard similar seasonalities,
see \eg  \cite{LefolMercier 98,DGMOP 01}.

However, as shown in Fig. 4,
the survival probability $\Psi(\tau)$ can be
fitted by a {\it stretched exponential} function:
$\exp \l[-(\tau/\tau_{0})^{\beta}/\Gamma(1+\beta)\r]$, with $\tau_0=6.6 s$
and $\beta=0.7$. The reduced chi-square of the fit is 0.71.

In a previous work
on bond futures \cite{MRGS 00},
according to theoretical
considerations on the properties of continuous-time random walks,
we suggested the Mittag-Leffler function
with a {\it power-law decay}  
as a suitable fit
for the empirical survival probability.
The above result does not contradict our previous findings.
In fact,
whereas for bonds futures we found waiting times greater
than $10,000 s\,,$
here we have only waiting times smaller than $200 s$,
 and the Mittag-Leffler function
is well approximated by the {\it stretched exponential}
as $\tau$   is small enough, see Eq. (1.4).

\begin{table}
\begin{tabular}{|l|l||l|l|l|}
\hline
\multicolumn{2}{|c||}{} & \multicolumn{3}{|c|}{$\tau_n$}\\
\cline{3-5}
\multicolumn{2}{|c||}{} &$0 \div 10 $ & $ 10 \div 20$ & $>20$\\
\hline
\hline
& $< -0.002 $ & 25 (38.9)& 21 (10.1)& 9 (6.0)\\
\cline{2-5}
& $-0.002 \div -0.001 $ & 516 (613.6)& 230 (159.5)& 122 (94.9)\\
\cline{2-5}
$\xi_n$ & $ -0.001 \div 0$ & 6641 (7114.3)& 2085 (1849.1) & 1338 (1100.6)\\
\cline{2-5}
& $ 0 \div 0.001$ & 31661 (31008.0) & 7683 (8059.2)& 4520 (4797.0)\\
\cline{2-5}
& $ 0.001 \div 0.002$ & 398 (464.4)& 179 (120.7)& 80 (71.9)\\
\cline{2-5}
& $ > 0.002$ & 34 (36.1) & 10 (9.4)& 7 (5.6)\\
\hline
\end{tabular}
\\
\caption{Contingency table between
log-returns $\xi_n$ and waiting times $\tau_n$.
Every cell contains the frequency observed
within the values considered and (in brackets) the theoretical
frequency which can be computed under the null hypothesis of
independence between $\xi_n$ and $\tau_n$.}
\end{table}

\begin{figure}[p]
\centering
 \centerline{\psfig{file=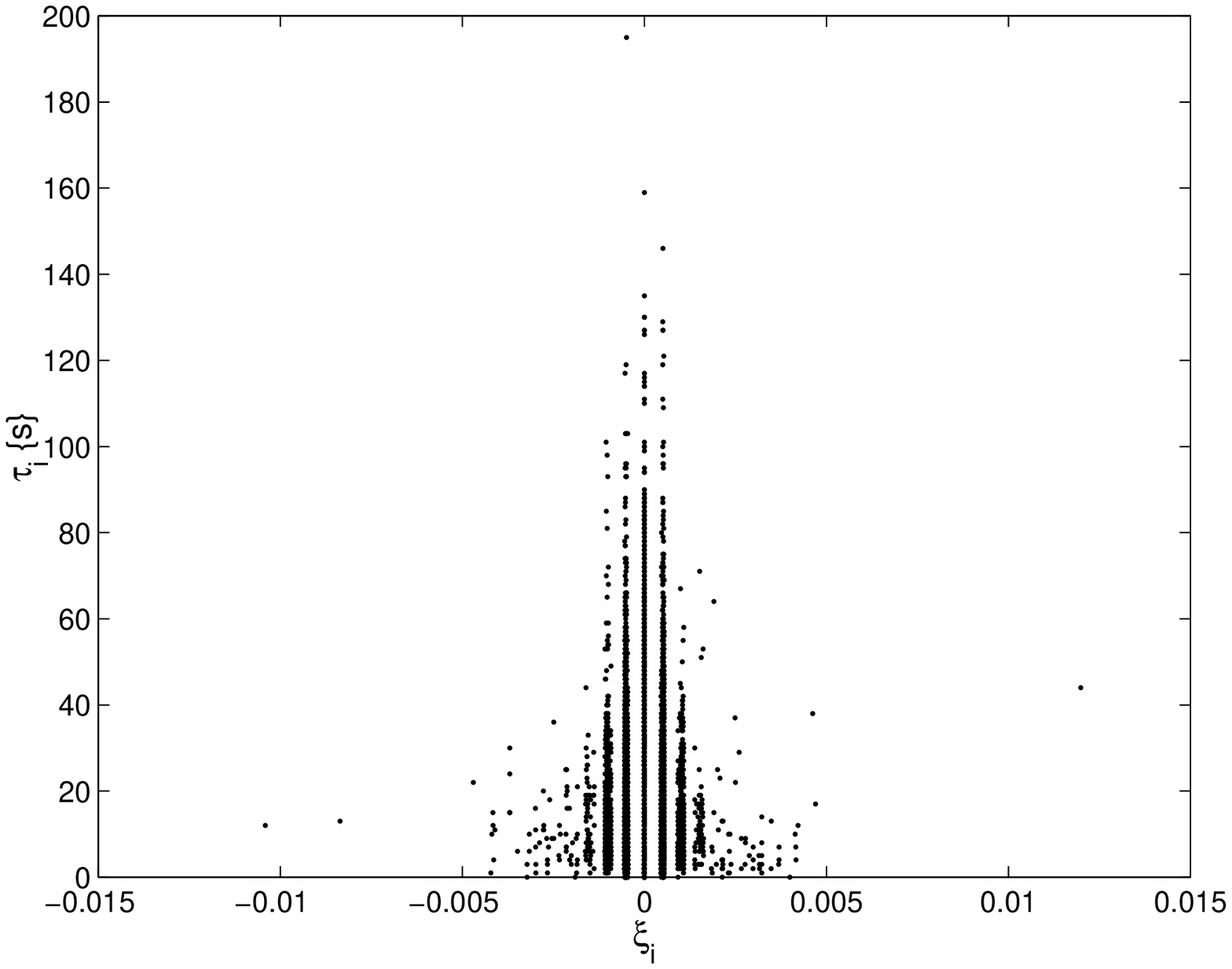,height=7.5truecm,width=10.0truecm}}
\caption{Scatter plot of waiting times $\tau_n$ as a function of the
corresponding log-returns $\xi_n$.}
\end{figure}

\begin{figure}[p]
\centering
 \centerline{\psfig{file=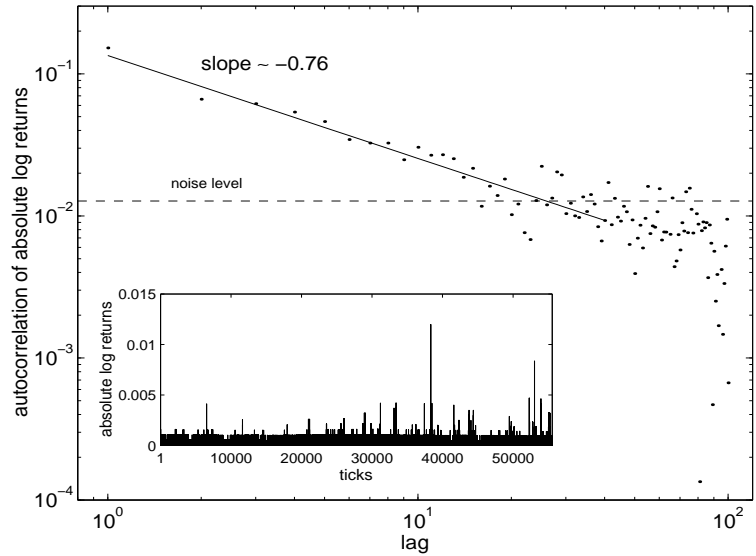,height=7.5truecm,width=10.0truecm}}
\caption{Autocorrelation function for the absolute value of
log-returns $\xi_n$. The inset shows the time series.}
\end{figure}

\begin{figure}[p]
\centering
 \centerline{\psfig{file=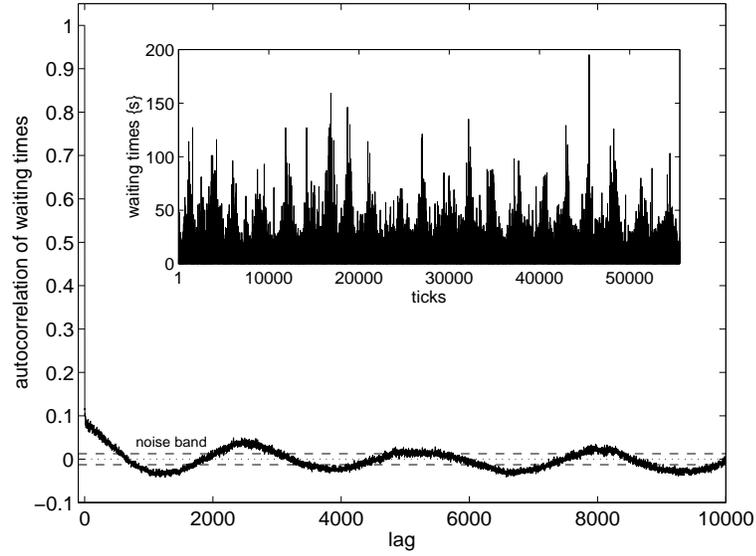,height=7.5truecm,width=10.0truecm}}
\caption{Autocorrelation function for the waiting times $\tau_n$.
The inset shows the time series.}
\end{figure}

\begin{figure}[p]
\centering
\centerline{\psfig{file=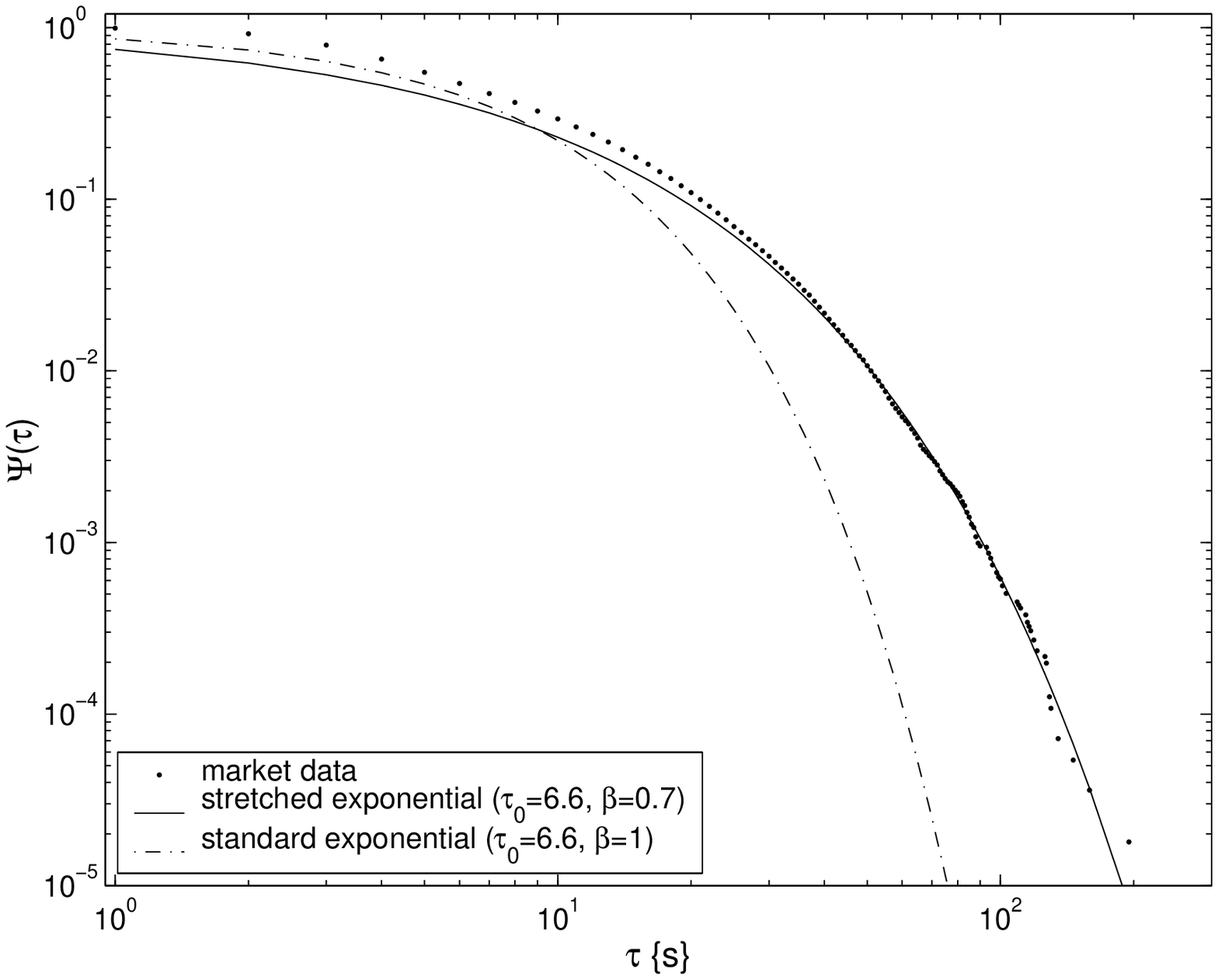,height=7.5truecm,width=10.0truecm}}
\caption{Survival probability. The stretched exponential (solid line)
is compared with the standard exponential (dash-dotted line).}
\end{figure}

\section{Summary}

A preliminary study of General Electric
high-frequency stock prices has been performed.
Some statistical properties of the log-return
and waiting-time random variables have been presented.
This study was inspired by previous theoretical and empirical work,
based on the phenomenological CTRW model of financial markets.

The main results are as follows:
the two random variables cannot be considered independent from each other;
the autocorrelation of log-returns absolute values exhibits a
power-law decay and reaches the noise level after about 250 s;
the autocorrelation of waiting times shows a 1-day periodicity,
corresponding to the daily stock market activity.




\end{document}